\documentclass[aps,showpacs,showkeys,preprint,prb,amsmath,amssymb,superscriptaddress]{revtex4}

\usepackage[english]{babel}
\usepackage[latin1]{inputenc}
\usepackage{color}
\usepackage{graphicx}
\usepackage{graphics}
\usepackage{setspace}
\begin{document}
\title{Phase transition in two-dimensional magnetic systems with dipolar interactions}
\author{L.A.S. M\'ol}
 \email{lucasmol@ufv.br}
 \affiliation{Departamento de F\'isica, Universidade Federal de Vi\c{c}osa, 36570-000, Vi\c{c}osa, Minas Gerais, Brazil}
\author{B.V. Costa}
 \email{bvc@fisica.ufmg.br}
\affiliation{ Departamento de F\'isica, Laborat\'orio de Simula\c{c}\~ao, ICEX,
UFMG, 30123-970, Belo Horizonte, MG, Brazil }
\date{\today}

\begin{abstract}
In this work we have used extensive Monte Carlo calculations to study the planar to paramagnetic phase transition in the
two-dimensional anisotropic Heisenberg model with dipolar interactions (AHd) considering
the true long-range character of the dipolar interactions by means of the Ewald summation. Our results are consistent
with an order-disorder phase transition with unusual critical exponents in agreement with our
previous results for the Planar Rotator
model with dipolar interactions. Nevertheless, our results disagrees with
the Renormalization Group results of Maier and Schwabl [PRB, {\bf 70}, 134430 (2004)] and
the results of Rapini {\it et. al.} [PRB, {\bf 75}, 014425 (2007)], where the AHd
was studied using a cut-off in the evaluation of the dipolar interactions. We argue that
besides the long-range character of dipolar interactions their anisotropic character
may have a deeper effect in the system
than previously believed. Besides, our results shows that the use of a cut-off radius in the
evaluation of dipolar interactions must be avoided when analyzing the critical behavior of
magnetic systems, since it may lead to erroneous results.

\end{abstract}

\maketitle
%
%

\section{\label{introd}Introduction}
Magnetism is one of the most studied subjects in physics.
In particular, the study of the statistical behavior of spin models
was the trend in the years 1960's and 1970's. They gave fruitful
contribution to the understanding of several phenomena, not only in
Physics, but in several other fields. The concept of symmetry breaking
and universality transcend the subject of physics to many other areas, being thus of great importance.
More recently, mainly due to the growing interest in magnetic thin-films, magnetic
nano-dots and arrays of magnetic nanoparticles
~\cite{Berger,bader,pappas,allenspach,mol_SI2,mol_SI,birsan,henning,estevez,estevez2,toscano11_1,toscano11_2,
carvalho-santos10,apolonio2009},
there were a renewed interest in the study of
spin models.

In many  magnetic systems a long
range dipole-dipole energy term has to be considered beside the exchange
interaction between neighboring sites and the anisotropies present in the system.
The study of such models is mainly associated with the development of
magnetic-nonmagnetic multilayer for the purpose of giant magnetoresistence
applications and magnetic nanoparticle arrays as an alternative for new media storage.
In addition, experiments on epitaxial magnetic layers have shown that a huge
variety of complex structures can develop in the system\cite{debell,bader,allenspach}. Rich magnetic
domain structures like stripes, chevrons, labyrinths, and bubbles associated
with the competition between dipolar long-range interactions and a strong
anisotropy perpendicular to the plane of the film were observed experimentally.
A lot of theoretical work has been done on the morphology and stability of
these magnetic structures\cite{,toscano11_1,toscano11_2,
carvalho-santos10,apolonio2009,debell,santamaria,carubelli,macisaac,whitehead,kwon}.
Beside that, it has been observed the existence
of a switching transition from perpendicular to in-plane ordering at low but
finite temperature\cite{pappas,allenspach}: at low temperature the film magnetization is perpendicular
to the film surface; rising temperature the magnetization flips to an
in-plane configuration. Eventually the out-of-plane and the in-plane
magnetization become zero. Although the structures developed in the system
are well known, the phase diagram of the model is still not completely understood.
%
%
\begin{figure}
\begin{center}
\includegraphics[scale=0.25]{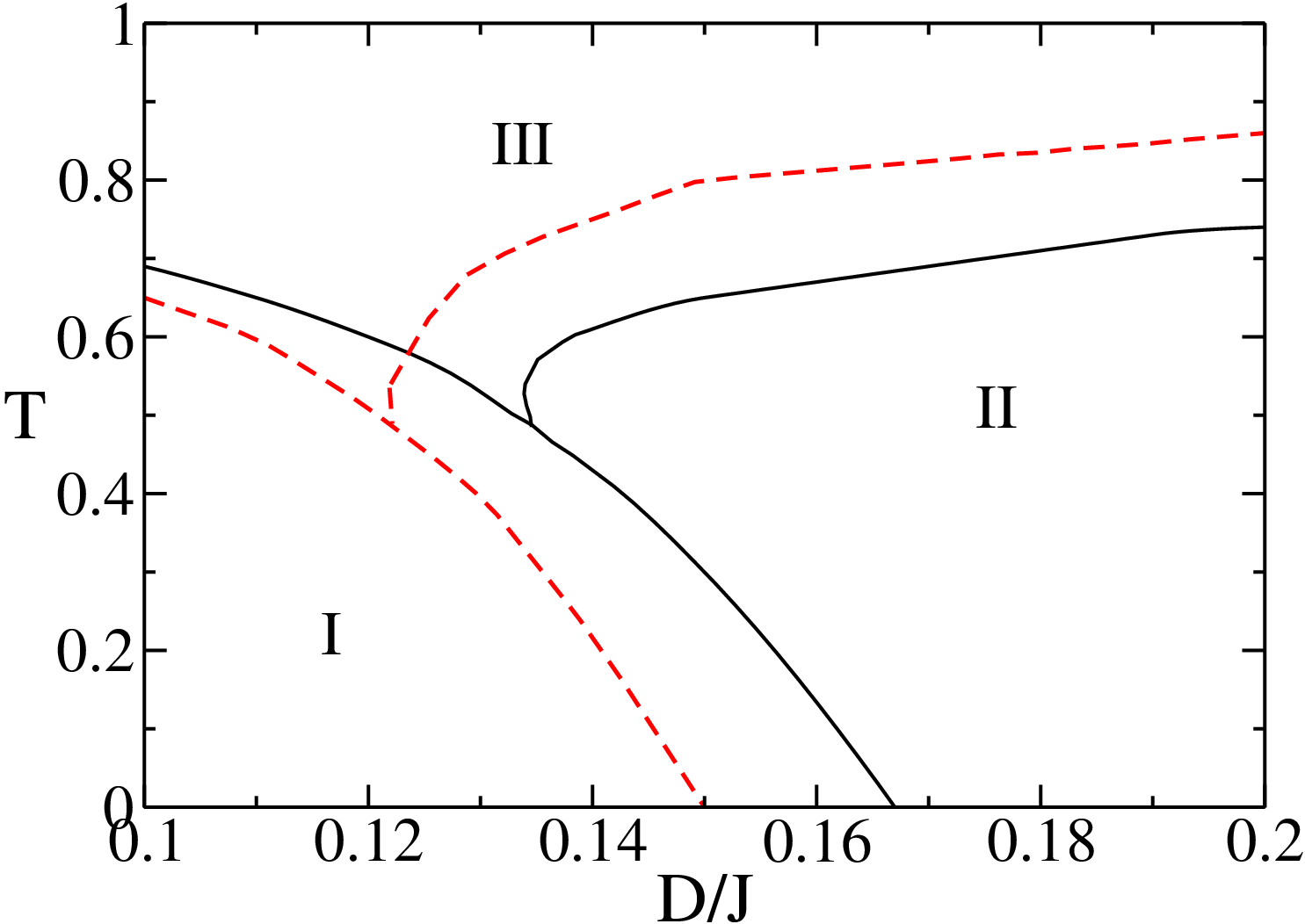}
\caption{\label{diagram} (Color online) Phase diagram of the anisotropic Heisenberg model with dipolar
interactions (AHd) for fixed $A/J=1$ in the $(D/J,T)$ space. The black solid line
represents the transition lines as obtained using a cut-off in the dipolar interactions
\cite{marcella} and the solid black line are the results obtained when full long-range interactions
are considered by means of the Ewald summation (this work). The phase I is an Ising-like phase characterized
by an ordered out-of-plane alignment of spins (that may present stripe-like configurations for full
long-rang interactions). Phase II
is an ordered planar ferromagnetic state and phase III is a paramagnetic one.}
\end{center}
\end{figure}
%

Here, our interest is in magnetic thin-films with an out-of-plane anisotropy as
described above. For such a system we can write a model Hamiltonian as follows
\begin{equation}\label{dipolo}
 H = -J \sum_{<i,j>}(\vec{S}_i\cdot \vec{S}_j) -A \sum_i (S^z_i)^2 +D \sum_{i \neq j} \left[  \frac {\vec{S}_i \cdot \vec{S}_j}{r^3_{i,j}}
 - \frac {\left( \vec{S}_i \cdot \vec{r}_{i,j}\right)\left( \vec{S}_j \cdot \vec{r}_{i,j}\right)}{r^5_{i,j}} \right],
\end{equation}
known as anisotropic Heisenberg model with dipolar interactions (AHd).
Here, we consider a ferromagnetic system, so that $J > 0$, $A$ is an easy-axis anisotropy,
and $D$ is the strength of the dipolar interactions. $\vec{r}_{i,j}$ is a vector connecting
sites $i$ and $j$ while $<i, j >$ means that the first
summation is to be evaluated for nearest neighbors only. For the dipolar
interactions the summation is evaluated over all pairs $i \neq j$ and in the single-ion anisotropy term
the summation is evaluated for all sites in the lattice.
For $D \neq 0$ the system is frustrated due to the
competition between the dipolar and the anisotropic terms. For small $D/J$
compared to $A/J$ we can expect the system to have an Ising-like behavior 
since an out-of-plane configuration of the spins is expected.
If $D$ is not too small we can expect a transition of the spins from
out-of-plane to in-plane configuration. For large enough $D$ out-of-plane
configurations become unstable such that, the system lowers its energy by
turning the spins into an in-plane ferromagnetic arrangement.
Earlier works on this
model, which discuss the phase diagram, were mostly done using
renormalization group approach and numerical Monte Carlo simulation
\cite{debell,maleev,bruno,taylor,whitehead,kwon,marcella,carubelli,santamaria,macisaac}.
They agree between themselves in the main features. The phase diagram
for fixed $A/J$ is schematically shown in Fig.\ref{diagram} in the
$(D/J,T)$ space. From Monte Carlo (MC) results it is found that there are
three regions labeled in Fig. \ref{diagram} as I, II, and III. Phase I corresponds
to an out-of-plane magnetization, phase II has in-plane magnetization,
and phase III is paramagnetic. The border line between phase I and phase
II is believed to be of first order and that between regions I and II is a second order one\cite{santamaria}.
The transition line between regions II and III has a non clear character.
Some authors reported they found a second order line\cite{santamaria}. In reference \onlinecite{marcella}
the author's claim that the transition is of the $BKT$ type. In many cases
a second order phase transition can be confused with a $BKT$ transition.
This can be due to the analysis of the finite size effects to be done in
not large enough lattices. Although the different results point out in the
direction of a second order or a $BKT$ transition~\cite{santamaria,marcella} between region II and III,
much care has to be taken because they were obtained by using a cutoff radius
$r_c$ in the dipolar interaction. The long-range character of the potential
is lost and then long-range order may not develop\cite{maleev}. As a consequence,
it will not be surprising if a completely different
scenario emerges when considering the true long-range dipolar interaction, since long-range
order is expected to be present\cite{maleev}.
In a recent paper, Maier and Schwabl\cite{maier} analyzed the phase transition in the
dipolar planar rotator (dPR) model via renormalization group techniques. It is expected
that the dPR model describes the critical behavior of the planar to paramagnetic
transition in the anisotropic Heisenberg model with dipolar interactions (AHd) since they
have the same simetry.
Their results indicate that the dPR model belongs to a new universality
class characterized by an exponential behavior of the magnetization,
susceptibility and correlation length. Besides that, the specific heat was
found to be non-divergent, as occurs in the BKT phase transition. More
recently M\'ol and Costa\cite{mol,mol2} studied, by using Monte Carlo simulations,
the AHd model in a bilayer system and the dPR model. In both
cases they found strong evidences for the transition between region II and III
to be in another universality class characterized by a mixed behavior between
order disorder and BKT transitions.
Besides, even in a 2D dipole lattice the scenario is not clear due to some
conflicting results (see, for example, references~\cite{baek,carbognani,rastelli,fernandez}).
It should be noticed however that in most of these works the full long-range character
of dipolar interactions are not taken into account properly. Indeed, the dipolar interaction
is conditionally convergent in two dimensions due to its anisotropic character, such that the use
of the minimal-image convention or the introduction of cut-offs in the potential may not describe the
real behavior of the system.

In this work we have used extensive
Monte Carlo simulations to study the transition line between
regions II and III (the planar to paramagnetic phase transition) in the
easy-axis Anisotropic Heisenberg model with dipolar interactions (AHd)
(see Eqn.\ref{dipolo}). Although the system in consideration has a significant
out-of-plane anisotropy our main interest is in a easy-plane symmetry brought about
by the dipolar interactions, so that for a suitable choice of the model parameters the spins
lie in the film plane. Indeed, our goal is to compare the results obtained for
the above mentioned model when full long-range dipolar interactions are considered by using the Ewald summation~\cite{weis,wang} (this work)
with those where a cut-off radius is introduced in the evaluation of dipolar
interactions\cite{marcella}. Our results clearly
indicate that considering the long-range character of dipolar interactions
the transition line between the planar and paramagnetic phases is of the
order-disorder type, being characterized
by a non-divergent specific heat and unusual critical exponents, instead of being a Berezinskii-Kosterlitz-Thouless
one as reported in reference \onlinecite{marcella}. This observation can be
of great importance since the effects of dipolar interactions is enhanced in
systems in the sub-micrometer scale, as magnetic nano-dots\cite{cowburn,julio,carvalho-santos10,apolonio2009}
and arrays of magnetic nanoparticles\cite{wang_SI,mol_SI2,mol_SI} and can shed extra light 
on some controversial results in the literature~\cite{baek,carbognani,rastelli,fernandez}.

\section{Simulation background}
In this work we followed the same methodology used in Refs.~\onlinecite{mol,mol2}.
The reader is referred to these works for a more detailed description, specially 
about the determination of critical temperature and exponents.
Our Monte Carlo procedure consists of a simple Metropolis algorithm\cite{livro_landau}
where one Monte Carlo step (MCS) consists of an attempt to assign a
new random direction to each spin in the lattice. To equilibrate the
system we have used $100 \times L^2$ MCS which has been found to be
sufficient to reach equilibrium, even in the vicinity of the transition.
In our scheme, two sets of simulations have been performed. In the
first one, we preliminarily explored the thermodynamic behavior of the
model in order to estimate the position of the maxima of the specific heat
and susceptibilities and the crossings of the fourth order Binder's cumulant.
In this first approach we used lattice sizes in the interval $20 \leq L \leq 50$.
Once the possible transition temperature is determined, we refined the
results by using single and multiple histogram methods. We produced the
histograms for each lattice size in the interval $20 \leq L \leq 120$ and
they were built at/close to the estimated critical temperatures corresponding
to the maxima and/or crossing points obtained in step 1.
To construct the histograms at least $2 \times 10^7$ configurations
were obtained using at least 3 distinct runs. These histograms are
summed so that we obtain a new histogram that allow us to explore
a wider range of temperature (an example of the use of histograms can be
found in Ref. \onlinecite{mol}). Periodic boundary
conditions are assumed in the directions $x$ and $y$. To take into account
the long range character of the dipolar interaction we use the Ewald summation
to calculate the energy of the system\cite{weis,wang}.

All simulations where done using a square lattice, $A/J=1$ and $D=0.3J$.
Energy was measured in units of $J$ and temperature in units of $J/k_B$, where
$k_B$ is the Boltzmann constant.
Our choice of $D=0.3J$ was to guarantee that the planar behavior of
the system was not much affected by the frustration existent 
near the multicritical point where the three lines shown in Fig. \ref{diagram}
come together.
We have devoted our efforts to determine a number of thermodynamic quantities,
namely the specific heat, magnetization, susceptibility, fourth order Binder's
cumulant and moments of magnetization as described elsewhere~\cite{mol,mol2}.

\section{Simulation Results}
Concerning the systems' magnetization no significant
size dependence is observed in low temperatures, unlike the results shown in
Ref.~\onlinecite{marcella} where a cut-off radius were used in the evaluation
of dipolar interactions. This may be an evidence that as the full long-range 
character of dipolar interactions are taken into account long-range order
develops, as expected by the results of Maleev\cite{maleev}.

\begin{figure}
\begin{center}
\includegraphics[scale=0.25]{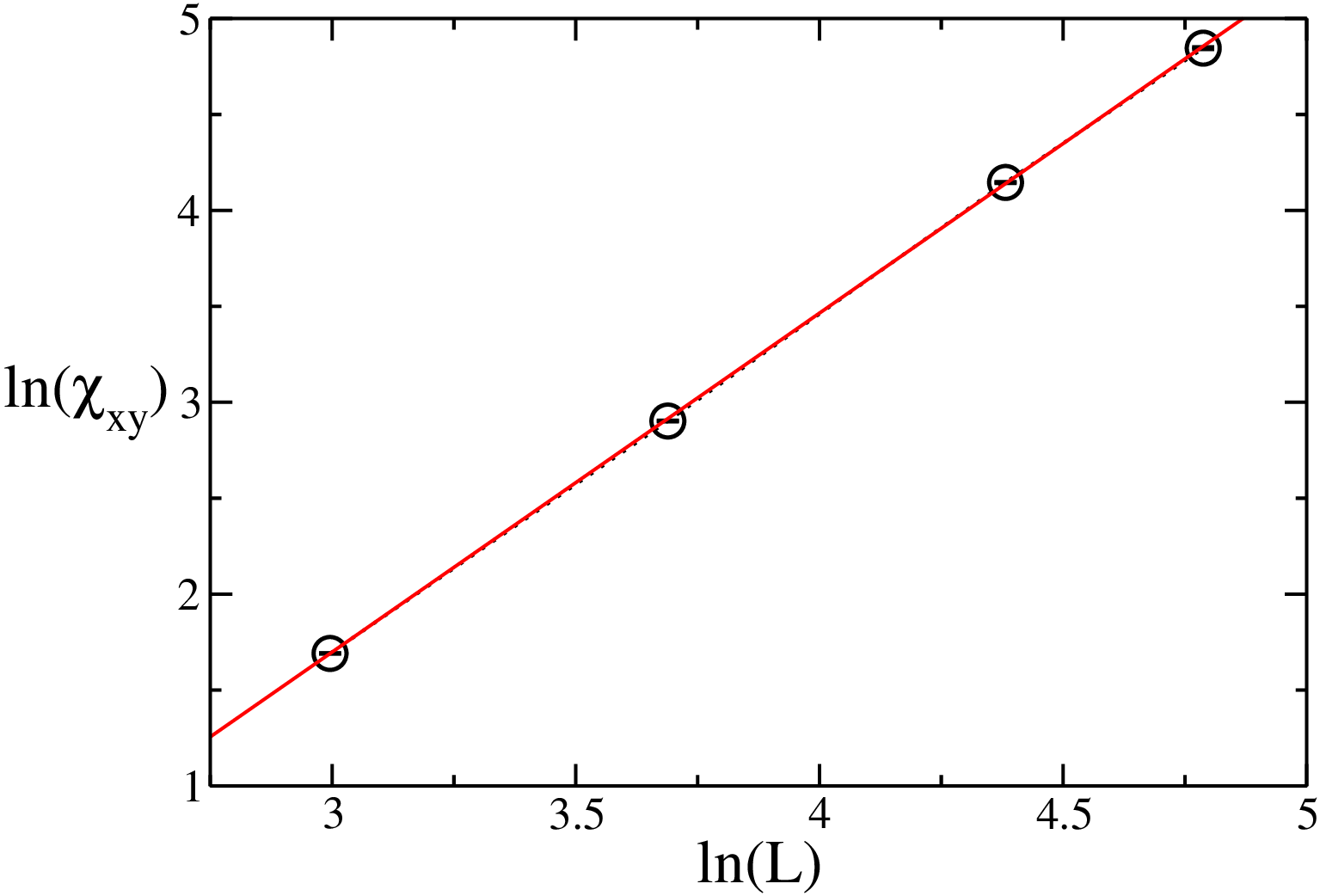}
\caption{\label{fss_sus} (Color online) Log-log plot of the maxima of the planar
susceptibility as a function of the lattice size. The solid red line shows the
best linear fit of the data given the exponent $\gamma /\nu=1.763(1)$.
The error bars are shown inside the symbols.}
\end{center}
\end{figure}

In figure~\ref{fss_sus} we show a $log-log$
plot of the maxima of the susceptibility as a function of the lattice
size for $L = 20, 40, 80$ and $120$. The data are very well adjusted
by a straight line with slope $\gamma/ \nu = 1.763(1)$ exhibiting
a power law behavior. This value of the exponent $\gamma /\nu$ is quite near
the expected one for a transition in the Ising universality class (1.75). 
Considering the Ising universality class we were able to determine the critical
temperature by using the location of the maxima of the specific heat and
susceptibility and the crossing point of the Binder's cumulant, which gives
$T_c^{Ising}=0.946(1)$. By using this value and plotting $\ln (M_{XY} \times \ln(L)$
at $T=T_c$ we have found $\beta/\nu==0.163(6)$, which is quite different from
the expected value for the Ising universality class ($0.125$).
%
%
%

The last result may indicate that the assumption of the Ising universality class may not be correct.
Indeed, by analyzing the moments of magnetization defined in Ref.~\onlinecite{mol}, we obtain
$1/\nu=0.82(2)$ and $T_c^{V_j}=0.943(1)$. This value
of the exponent $\nu$ contrasts with the expected for the Ising universality class, although 
the value for the critical temperature is approximately the same. Reanalyzing our previous 
estimates for the critical temperature obtained using the location of the specific heat
and maxima of susceptibilities using this new value of the exponent $\nu$ we obtain:
$T_c^{c_v}=0.945(1)$ and $T_c^{\chi}=0.943(1)$ (it is worthy to note that in the analysis
of the specific heat data the point corresponding to $L=20$ was disregarded in both cases).
Looking to the crossing point of the Binder's
cumulant we have found $T_c^{U_4}=0.944(2)$.
We have thus, as our new estimate for the mean critical temperature $T_c=0.944(1)$. 
Using this new value of the critical temperature we obtain $\beta/\nu=0.149(7)$ in the
analysis of the magnetization data.

\begin{figure}
\begin{center}
\includegraphics[scale=0.25]{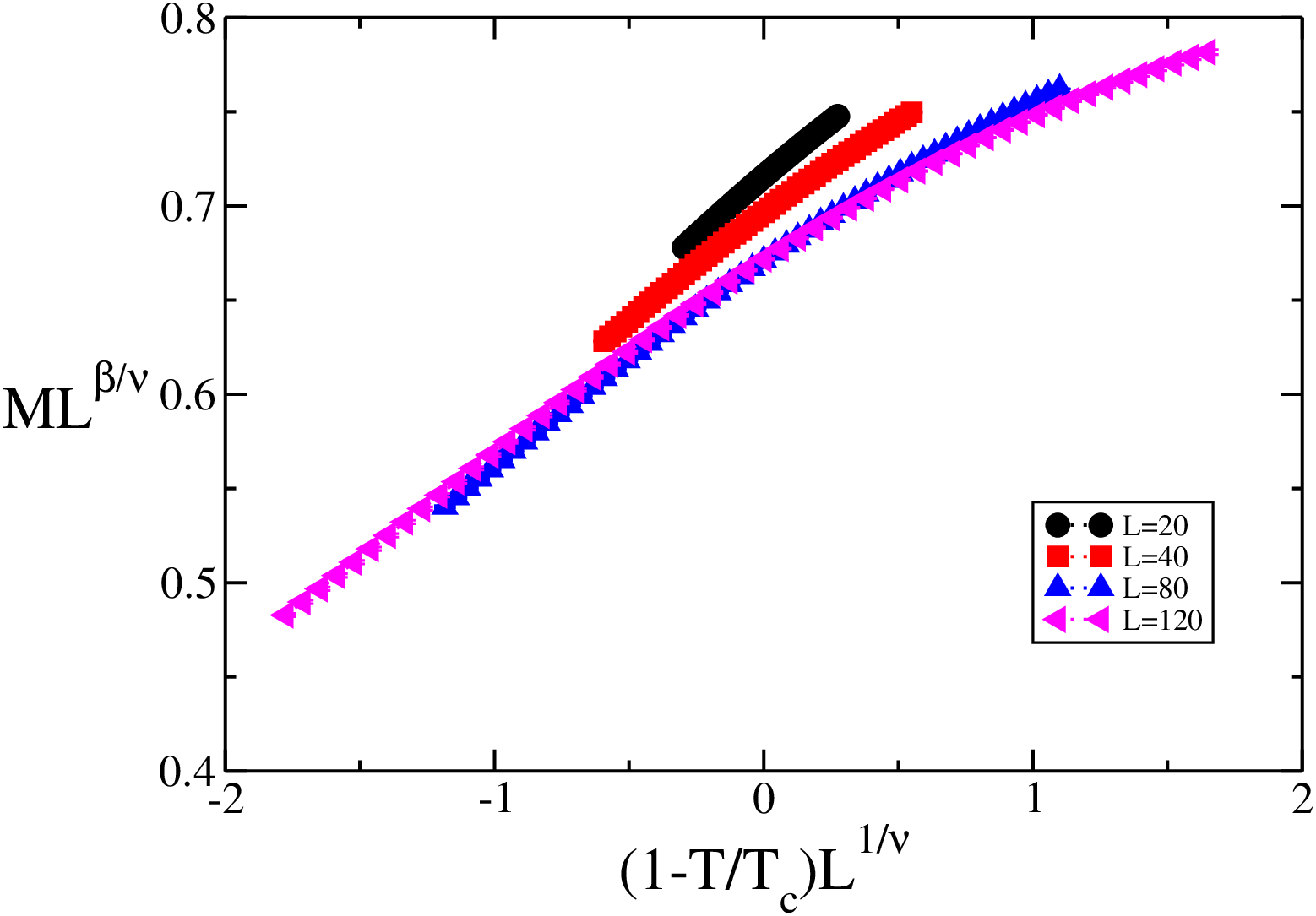} \\ \includegraphics[scale=0.25]{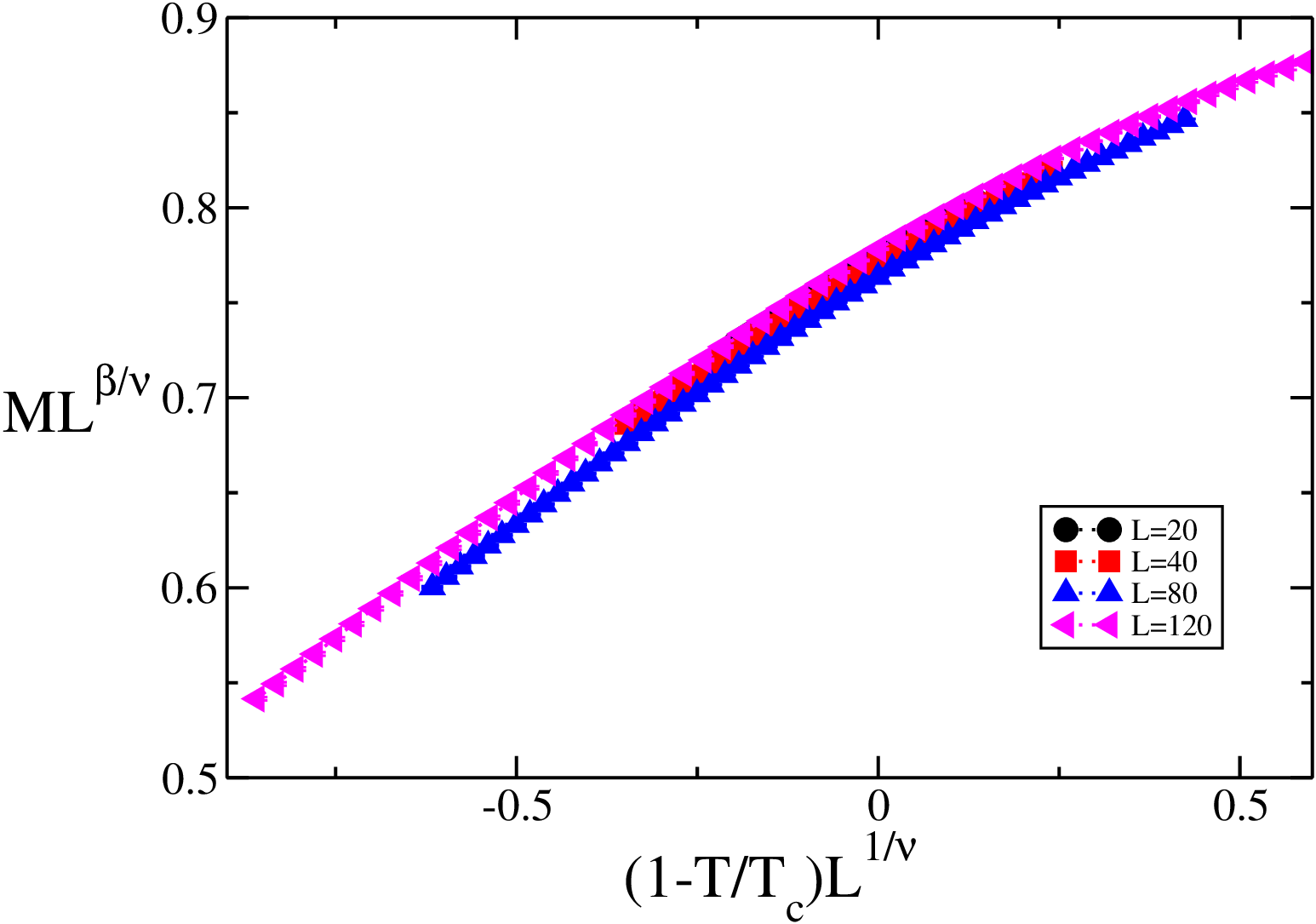}
\caption{\label{sc_mag} (Color online) Scaling plots of magnetization considering
the Ising-like behavior (top) and an order-disorder transition characterized
by the exponents shown in the last line of table~\ref{table} (bottom).
}
\end{center}
\end{figure}

To distinguish between these scenarios in figure~\ref{sc_mag} we show a
scaling plot of the magnetization obtained with the multiple histogram technique according to its finite size scaling function
($m \approx L^{\frac{\beta}{\nu}} \textbf{\textit{M}}\left( tL^{\frac{1}{\nu}} \right) $)
considering two possibilities: ($i$) the Ising-like behavior ($T_c^{Ising}=
0.946(1)$, $\nu=1$ and $\beta=0.125$) and ($ii$) an
order-disorder critical behavior with exponents $\nu=1.22(3)$ and $\beta=0.18(1)$ and critical
temperature $T_c=0.944(1)$. As can be seen, the scaling plot obtained
assuming the Ising universality class does not describe our data as good as the
results considering a new universality class. Besides, doing the same analysis with
susceptibility and Binder's cumulant no significant deviations were observed between
these two possibilities. Indeed, the values obtained in this study
are in good agreement with those obtained for the same model in a bilayer system\cite{mol}
and for the dipolar Planar Rotator model\cite{mol2}. To clarify, in table~\ref{table}
we show the exponents for the Ising model, the results obtained by Maier and Schwabl for the
dPR model, the results of Refs.~\onlinecite{mol,mol2} and the results of this work.

%
%
%
%
\begin{table}\begin{center}
\begin{tabular}{lccccccc}
\hline \hline
Model          & $T_c$         & $\nu$ & $\gamma$      & $\beta$       & $\alpha$      \\
\hline
Ising           & 2.269         & 1     & 1.75          & 0.125         &  0 ($\ln$)    \\
dPR (Maier)     &               &       & 1             & 1/2           & -2            \\
AHd (bilayer)   & 0.890(4)      & 1.22(9)& 2.1(2)       & 0.18(5)       &  -0.55(15)    \\
dPR            & 1.201(1)      & 1.277(2)& 2.218(5)    & 0.2065(4)     &  -1.1(1)      \\
AHd (this work)            & 0.944(1)      & 1.22(3)& 2.15(5)    & 0.18(1)      &  -0.44(18)    \\
\hline \hline \end{tabular} \caption{ \label{table} {In this table we show the critical temperature
and exponents for the 2D Ising model\cite{onsager} (first line), the results of Maier and Schwabl\cite{maier}
for the dPR model, the results of MC calculations in the bilayer AHd model with a cut-off in the interactions
\cite{mol}, the results of MC calculations for the dPR model\cite{mol2} and the results of this work. }} \end{center}
\end{table}

\begin{figure}
\begin{center}
\includegraphics[scale=0.25]{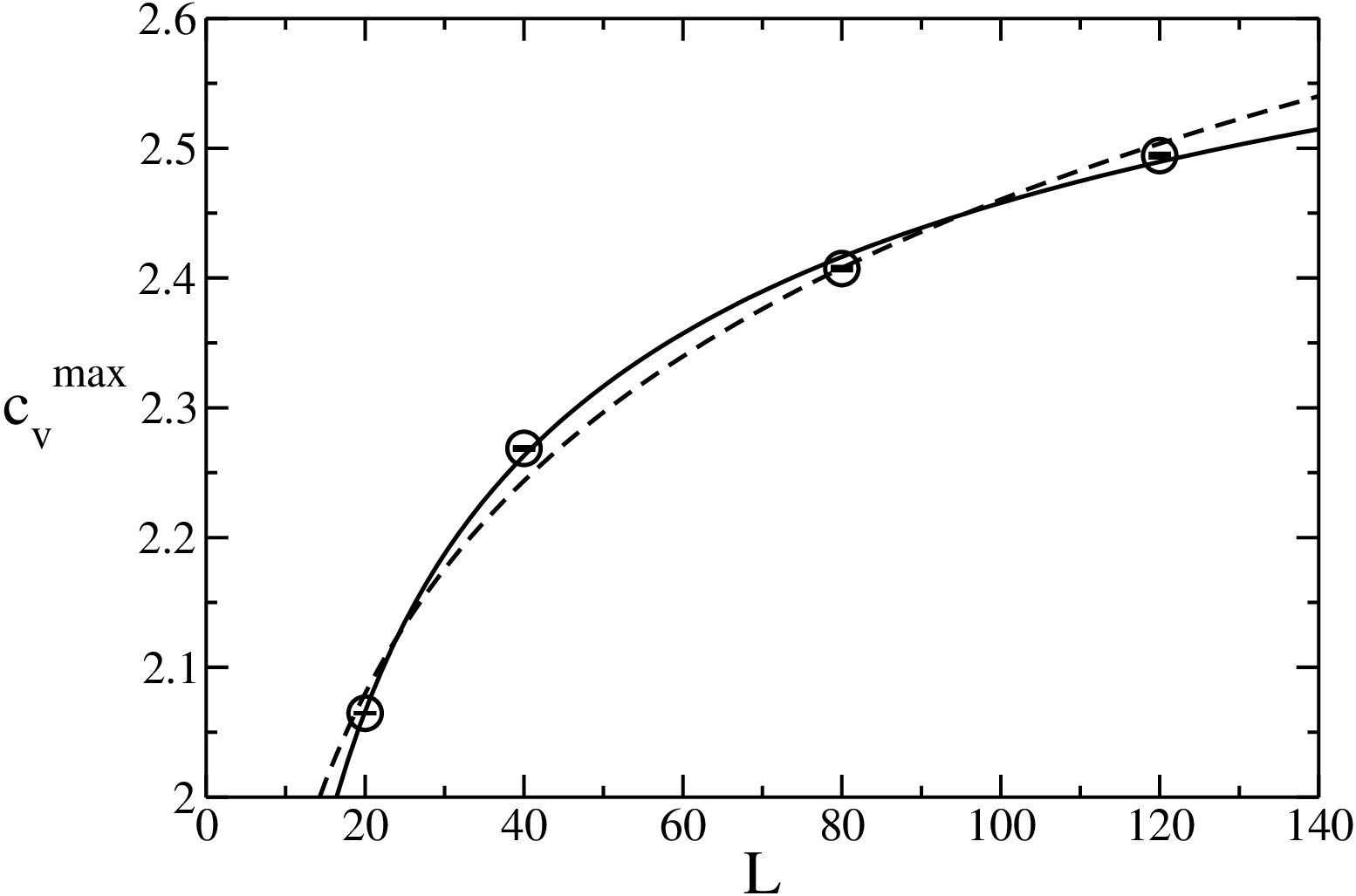}
\caption{\label{fss_cv} (Color online) Specific heat maxima as a function of the 
lattice size. The dashed line is the best non-linear fit considering an logarithmic 
divergence and the dashed line shows the best non-linear fit considering a non-divergent power law behavior.
}
\end{center}
\end{figure}

So far, everything corroborates to a order-disorder phase transition with non-conventional
critical exponents. However, the scale relations\cite{privman}
$\alpha + 2\beta + \gamma =2$ and $\nu d=2-\alpha$ are believe to be satisfied. Using the
values shown in table~\ref{table} and the first relation we should have $\alpha = -0.51(7)$ and using
the second relation $\alpha=-0.44(6)$ indicating the possibility that the specific heat
does not diverges. Indeed, to have an better agreement between the results of this work
and those of Refs.~\onlinecite{mol,mol2} the specific heat should be non-divergent.
As one knows, to distinguish between an logarithmic divergence or an slowly power law
divergence or even an non-divergent power law, many orders of magnitude are needed. 
Nevertheless, a careful analysis of the data could give us a clue. In figure~\ref{fss_cv}
we show our data for the maxima of the specific heat as a function of the lattice size
adjusted by two different methods. The dashed line represents the best fit of a logarithmic divergence
($a\ln (L)+b$), the solid line is for a non-divergent power law behavior ($-a L^{-b}+c$). As can be clearly seen, the
non-divergent power law describes better the data. Indeed, the $\chi^2/dof$ values obtained
are $4.7\times 10^{-4}$ for the logarithmic divergence and $1.4\times 10^{-6}$ for the non-divergent
power law. The value obtained for the exponent $\alpha/\nu$ from the adjust is $-0.36(14)$, and it
is also shown in table~\ref{table}, together with the results for the dPR and a bilayer AHd models.
%
%
%

\section{Discussion}
In this work we have studied the phase transition in the
anisotropic Heisenberg model with dipolar interactions (AHd).
Our main goal was to look for possible differences in the critical
behavior of the planar to paramagnetic phase transition when full long-range
dipolar interactions are considered or not. We have found that the use of the 
full long-range interaction by means of the Ewald summation~\cite{weis,wang}
lead to different results, i.e., while the introduction of a cut-off radius at
five lattice spacings lead to a BKT transition (Ref.~\onlinecite{marcella}),
the use of the Ewald summation lead to a order disorder transition with unusual
exponents and a non divergent specific heat. Indeed, it would be interesting
to present a detailed study of the effects in the critical behavior of the
system as the cut-off radius increases. Nevertheless, this study is
beyond the scope of this paper and will be addressed in a near future.
On the other hand some points deserves a more detailed discussion.

In what follows we present a discussion very similar to that done in Refs.~\onlinecite{mol,mol2}.
In the low temperature phase ($T < T_c$) the magnetization of the model
does not display any significant decrease as the lattice size is augmented.
That is a clear indication of an order-disorder transition. This fact
is in disagreement with the results obtained in the work of Rapini
et al.\cite{marcella} where was reported a BKT transition.
Our results are very well
described by a finite size scaling theory based on the existence of a
low temperature phase with long range order and finite correlation length\cite{privman}.
In a BKT phase transition there is no long-range order in the low temperature
phase, consistent with the Mermin Wagner theorem\cite{mermin-wagner}. Indeed, the results of
Maleev\cite{maleev} predict the existence of long-range order at low temperatures
in the AHd model. Our results are consistent with this scenario. As discussed
earlier, recent renormalization group results by Maier and Schwabl\cite{maier}
predicted that this system may belong to a new universality class,
characterized by the presence of long-range order at low temperatures
and by an exponential behavior of thermodynamic quantities in the
vicinity of the critical temperature. In the Maier and Schwabl scenario
the correlation length diverges as $ \chi \sim \exp (bt^{1/2})$
when the critical temperature ($T_c$) is approached from the high
temperature side, similar to the behavior of the BKT phase transition,
while the behavior of other thermodynamic quantities are given by powers
of the correlation length. Nevertheless, our results for the AHd model are
very well described by power law divergences of the thermodynamic quantities.
As can be seen in figure~\ref{sc_mag}, we have obtained a very good collapse of the curves
from different lattice sizes for the magnetization and similar results
were also found for the susceptibility and 
Binder's cumulant. These curves show that the critical exponents obtained
and the conventional finite size scaling theory, that assumes a power
law behavior of thermodynamic quantities, describe the Monte Carlo data
accurately, indicating that the phase transition in the AHd model is a
conventional order-disorder phenomenon with unusual critical exponents.
In order to definitely rule out the possibility of this phase transition
being in the new universality class proposed by Maier and Schwabl, we
should make a comparison of our Monte Carlo results, using a finite size
scaling theory based in their predictions and the conventional finite
size scaling theory used here. Unfortunately, it is not very clear in
the literature how to obtain a finite size scaling theory for exponential
divergences. Using a simple replacement of the correlation length by the
lattice size (in a manner similar to that made by Challa and Landau in reference~\cite{challa}),
which should be the first choice, does not give a good collapse of the curves,
mainly because the determination of the critical temperature is quite
imprecise in this case and the collapse of the curves depends appreciably
on the value used for the critical temperature. In any case, using values
for the critical temperature close to the maxima of the susceptibility we
were not able to obtain even a reasonable collapse of the curves. At this
point one can argue: Why do renormalization group results not agree with
Monte Carlo simulations? Actually, the RG study of Maier and Schwabl\cite{maier}
is based upon some approximations, for instance by using a continuous
version of the model. Since the dipolar interactions have an intrinsic anisotropy,
which depends in a complicated manner on the location of each spin in the lattice,
the lattice geometry could have a strong effect in the system. The identification
and a detailed discussion of the points of the RG study of the model that are
behind the discrepancy between our results is beyond the scope of this paper.

Although it was shown that the unusual exponents describes better
the data, specially for the magnetization, the transition may be in the Ising 
universality class as well, since corrections to scaling were not taken into account
and the lattice sizes used may not be large enough. Indeed, as can be seen in
figure~\ref{sc_mag} the use of the Ising universality class exponents describes
well the data for the largest lattices studied. Nevertheless, it does not seems
to be a good choice simply disregard the data for the lattices with $L=20$ and $40$, 
leaving only two lattice sizes to be analyzed. Thus it is more prudent to not completely
rule out the possibility of this phase transition to belong to the Ising universality class.
On the other hand, our previous results for the same model in a bilayer system~\cite{mol} and the results
for the dPR model~\cite{mol2} were also well described by the same critical behavior found
here, such that we still believe that this phase transition is more likely to belongs
to a new universality class with unusual exponents.
Studies in much larger lattices could remove this ambiguity, nevertheless
the computational time needed for such a study turns it impracticable at the moment.

Concerning the
origin of the order disorder transition the question is even more complicated.
The long-range order observed at low temperatures is expected to occur only when
full long-range interactions are present. Nevertheless, in a recent study of
the AHd in a bilayer system\cite{mol} using a cut-off in the
dipolar interaction, we found the same critical behavior. The same results were also
obtained in a study of the dipole planar rotator model in two dimensions\cite{mol2}
with the dipole long-range interaction treated by means of the Ewald summation.
In both cases the critical exponents agree quite well with those obtained in the
present work (see table~\ref{table}). This observation indicates that the anisotropic character of dipolar
interactions may be the main factor responsible for the observed critical phenomena.
Indeed, this observation is not new in the literature. As an example, Fernandez and
Alonso\cite{fernandez} stated that ``Anisotropy has a deeper effect on the ordering of systems
of classical dipoles in 2D than the range of dipolar interactions''. In this work
the authors found that the inclusion of a quadrupolar anisotropy drastically
changes the phase transition behavior of a system of classical dipoles. Apparently,
in our system the intrinsic anisotropy of dipolar interactions plays an essential
role in the determination of the universality class of the AHd model. The possible
new universality class is not surprising. In the theory of critical phenomena\cite{stanley}
it is expected that the critical exponents, and thus the universality classes,
depend only on the spatial dimensionality of the system, the symmetry and dimensionality
of the order parameter, and the range of the interactions within the system,
characteristics not shared by the AHd model and models of well known universality
classes.

As a final remark we would like to stress that these results are much important
when the critical behavior of magnetic
systems with dipolar interactions is being considered. They show that the use
of a cut-off radius in the evaluation of dipolar interactions may lead to erroneous
results. This study may be a guide for future works in what concerns the
introduction or not of a cut-off radius in the study of critical behavior of
magnetic systems with dipolar interactions (see for example Ref.~\onlinecite{mol_comment}).

{We would like to thank Professor D P Landau for helpful discussions.
Numerical calculation was done on the Linux cluster at Laborat\'orio
de Simula\c{c}\~ao at Departamento de F\'{\i}sica UFMG. We are grateful
to CNPq and Fapemig (Brazilian agencies) for financial support.}
\bibliography{AHd_v1}

\end{document}